\documentclass[conference]{IEEEtran}
\IEEEoverridecommandlockouts
\usepackage{cite}
\usepackage{amsmath,amssymb,amsfonts}
\title{Aligned mathematics}
\usepackage{algorithm} 
\usepackage{algpseudocode} 
\usepackage{graphicx}
\usepackage{textcomp}
\usepackage{xcolor}

\def\BibTeX{{\rm B\kern-.05em{\sc i\kern-.025em b}\kern-.08em
    T\kern-.1667em\lower.7ex\hbox{E}\kern-.125emX}}

\begin{document}

\title{Intelligent Resource Allocation for UAV-Based Cognitive NOMA Networks: An Active Inference Approach}

\author{felix.obite }
\date{August 2022}

\author{\IEEEauthorblockN{Felix~Obite\textsuperscript{1,2}, Ali~Krayani\textsuperscript{1}, Atm~S.~Alam\textsuperscript{2}, Lucio~Marcenaro\textsuperscript{1}, Arumugam~Nallanathan\textsuperscript{2}, Carlo~Regazzoni\textsuperscript{1}}
\IEEEauthorblockA{\textsuperscript{1}\text{DITEN, University of Genova, Italy} \\
\textsuperscript {2}\text{EECS, Queen Mary University of London, United Kingdom}\\
\small emails:felix.obite@edu.unige.it, ali.krayani@ieee.org \{lucio.marcenaro, carlo.regazzoni\}@unige.it,\                                     \{a.alam, a.nallanathan\}@qmul.ac.uk }
}

\maketitle

\begin{abstract}
Future wireless networks will need to improve adaptive resource allocation and decision-making to handle the increasing number of intelligent devices. Unmanned aerial vehicles (UAVs) are being explored for their potential in real-time decision-making. Moreover, cognitive non-orthogonal multiple access (Cognitive-NOMA) is envisioned as a remedy to address spectrum scarcity and enable massive connectivity. This paper investigates the design of joint subchannel and power allocation in an uplink UAV-based cognitive NOMA network. We aim to maximize the cumulative sum rate by jointly optimizing the subchannel and power allocation based on the UAV's mobility at each time step. This is often formulated as an optimization problem with random variables. However, conventional optimization algorithms normally introduce significant complexity, and machine learning methods often rely on large but partially representative datasets to build solution models, assuming stationary testing data. Consequently, inference strategies for non stationary events are often overlooked. In this study, we introduce a novel active inference-based learning approach, rooted in cognitive neuroscience, to solve this complex problem. The framework involves creating a training dataset using random or iterative methods to find suboptimal resource allocations. This dataset trains a mobile UAV offline, enabling it to learn a generative model of discrete subchannels and continuous power allocation. The UAV then uses this model for online inference. The method incrementally derives new generative models from training data by identifying dynamic equilibrium conditions between required actions and variables, represented within a unique dynamic Bayesian network. The proposed approach is validated through numerical simulations, showing efficient performance compared to suboptimal baseline schemes.
\end{abstract}


\section{Introduction}
As the world advances towards the sixth generation (6G) wireless networks, which demand complete autonomy, numerous applications now necessitate the use of Unmanned Aerial Vehicles (UAVs) in complex and dynamic surroundings. In these scenarios, UAVs must rely solely on their onboard sensors to comprehend the environment they navigate through and efficiently accomplish their objectives \cite{elmokadem2021towards}. UAVs are utilized as flying aerial stations to serve ground users, facilitating data access, extending coverage, and enhancing communication rates \cite{9120492}. In contrast to conventional terrestrial wireless communication infrastructures, UAVs possess the capability to adapt their positions dynamically, ensuring they maintain accurate channel conditions. Cognitive Radio (CR) technology is a promising solution to the problem of spectrum scarcity. The basic idea of the overlay CR paradigm is to efficiently utilize the available radio spectrum by allowing secondary users (SUs) to access and share the spectrum opportunistically when the primary users (PUs) are not using it. Due to its ability to offer superior spectral efficiency and support huge connectivity, non-orthogonal multiple access (NOMA) has been chosen as a novel technology to substantially enhance the throughput of wireless networks. This approach involves exploiting the power domain to perform superposition coding at the transmitter and utilizing successive interference cancellation (SIC) at the receiver to distinguish signals from multiple users \cite{he2019joint}. Hence, the concept of UAV-enabled Cognitive NOMA is perceived as a strong candidate to improve the performance of future wireless networks. 

Nevertheless, to fully harness the benefits promised by future networks, a major challenge is how to maximize the sum rate of mobile users through joint sub-channel and power allocation. 
Additionally, the UAV's trajectory significantly influences the channel gains of mobile users.
It is worth noting that optimal joint subchannel and power allocation in NOMA have been proven to be NP-hard \cite{salaun2018optimal}. 
The existing works have relied on the UAV's perfect knowledge of the positions of ground users to derive the channel gain explicitly from a particular radio propagation model, and to solve mobility problems using traditional optimization techniques or machine learning (ML) models. In practice, this assumption is often not realistic, because of their inherent variability and complexity. While traditional optimization schemes have demonstrated outstanding performance, they lack inherent adaptability and typically rely on fixed objectives \cite{9120492}.
On the other hand, deep learning (DL) requires a large amount of labeled data for training, which is challenging to generate in a complex radio environment. Reinforcement learning (RL) algorithms come with several limitations that can hinder their practical applicability. They often demand extensive interactions with the environment to learn effective policies, which can be impractical in real-world scenarios, requiring significant trial and error \cite{kurenkov2018reinforcement}. Also, RL models usually struggle to generalize knowledge from one environment to another. Learning in a specific scenario may not easily transfer to new and different situations \cite{ccatal2020learning}. 

Conversely, active inference, an emerging approach from theoretical neuroscience, offers a comprehensive brain theory that unifies action, perception, and inference for adaptive systems\cite{friston2017graphical}. Biological agents exhibit preferred states (or prior preferences) that impact their interactions with the external world and can update under random conditions \cite{sajid2022active}. 
Over the past few years, active inference has been employed in numerous applications, including decision-making in uncertain situations, structure learning, and navigation. An extensive summary of these applications can be found in \cite{da2020active}. 

In this paper, motivated by \cite{9829873}, we propose a new active inference-based learning method called Active Generalized Dynamic Bayesian Network (Active-GDBN) to intelligently maximize the sum rate of a UAV-based cognitive NOMA network. The main novel contribution of Active-GDBN, which sets it apart from traditional optimization techniques and ML models, lies in its adaptive or real-time belief updating. Active-GDBN allows an agent to continuously update its beliefs and actions based on incoming sensory inputs, which aligns with the way biological systems operate. This formulation is inspired by the general interpretation of adaptive behaviour, where the UAV can adjust its perception of the environment and preferences in response to new information, allowing it to allocate resources efficiently based on its mobility.

The remaining content in this paper is outlined as follows: Section  \ref{Sec_relatedWork} introduces the related work. Section \ref{Sec_systemModel_probForm} describes the system model and problem formulation. The proposed method for intelligent resource allocation is explained in Section \ref{Sec_proposedMethod}. Section \ref{Sec_simulationResults} presents the simulation results and analysis. Lastly, Section \ref{Sec_conclusion} concludes the paper.

\section{Related Work}
\label{Sec_relatedWork}
Most research efforts have focused on convex or non-convex optimization schemes for UAV trajectory design to maximize wireless network throughput \cite{wu2018joint,li2019fundamental,al2014optimal}. These efforts have been directed towards both stationary and flying UAVs. In the case of a static UAV scenario, the study \cite{al2014optimal} concentrated on optimizing the UAV's altitude to achieve the highest coverage probability for ground users. Meanwhile, in \cite{lyu2016placement}, authors optimized the positions of several UAV base stations to maximize the throughput of ground users. Conversely, in the flying UAV setup, a joint design of the UAV-relay's trajectory and power allocation was presented in \cite{zeng2016throughput} to optimize the end-to-end throughput. 

In addition to optimization-based approaches, in recent times, reinforcement learning (RL) has been applied to communication networks, especially in UAV networks \cite{bayerlein2018trajectory,challita2019interference,zeng2019path}. Typically, these studies utilize RL to address offline optimization tasks, training the UAV to follow a path repeatedly under stationary conditions. A recent study \cite{9120492} addressed a more complex scenario and proposed an enhanced RL algorithm that incorporates expert knowledge of the wireless channel to optimize a UAV's dynamic maneuver, aiming to maximize the sum rate of mobile users. However, the study in \cite{9120492} assumes specific CSI that might not generalize to scenarios with time-varying CSI.
Different from traditional optimization and existing studies (see \cite{da2020active}), where active inference is examined in the context of predefined discrete state spaces, this study considers a more complex and continuous dynamic scenario by jointly optimizing subchannels and power allocation, taking into account UAV's mobility.

\section{System Model and Problem Formulation}
\label{Sec_systemModel_probForm}
\subsection{System Model}
\begin{figure}[htp]
    \centering
    \includegraphics[width=8.0cm]{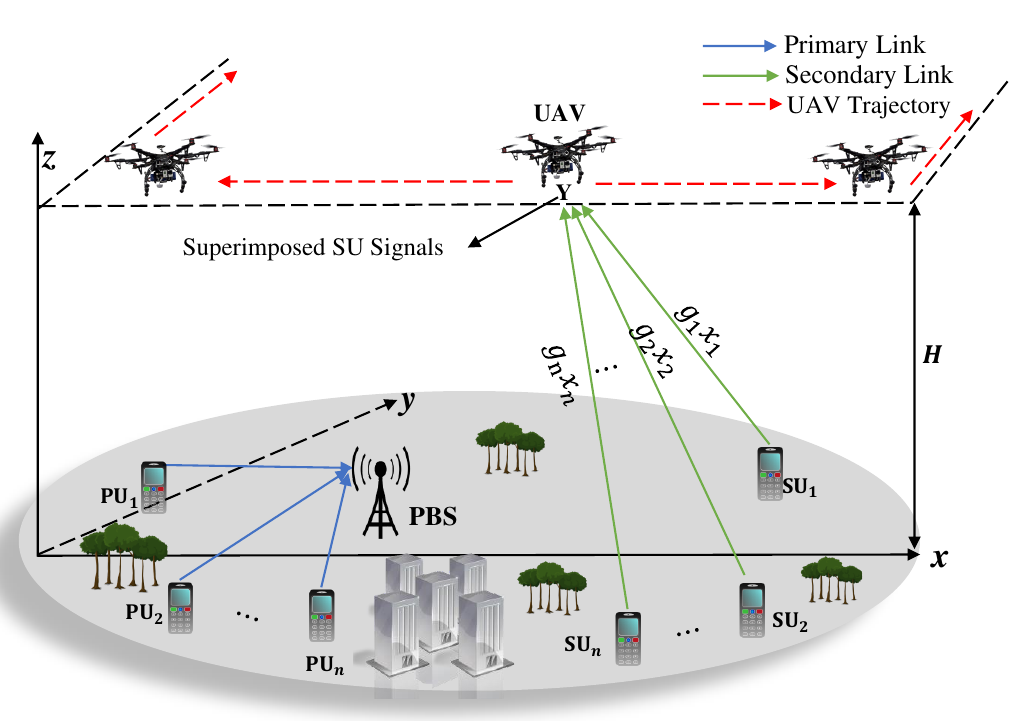}
    \caption{Illustration of system model with uplink NOMA.}
    \label{fig-system-model}
\end{figure}

As shown in Fig.~\ref{fig-system-model}, we investigate an uplink NOMA setup where a mobile UAV provides service to $N$ randomly moving SUs within a cell area. 
In practical applications, a single UAV communication is useful for emergency service recovery and delay-tolerant tasks like periodic data collection from ground sensors, which can be sufficient and cost-effective \cite{zeng2017energy}. There is a primary channel comprising a primary base station (PBS) that serves PUs.
The overall bandwidth is evenly divided into $K$ orthogonal subchannels, which eliminates interference between subchannels. The set of SUs and subchannels are denoted as $\mathcal{N}=\lbrace1,2,\cdots,N\rbrace$ and $\mathcal{K}=\lbrace1,2,\cdots,K\rbrace$, respectively.
At each time step, the positions of the mobile SUs are randomly updated within the 3D space. Each user moves independently and randomly in all three dimensions (x, y, and z) due to the random perturbation applied to their initial positions. The UAV's trajectory follows a straight line path based on randomly generated and normalized direction. The UAV's position is then updated based on this direction and the maximum velocity at each time step. We adopted the third-generation partnership project (3GPP) simulation standardization, where drones initial movement start from randomly chosen positions within the network. Subsequently, they travel at a constant speed and height, moving in straight lines while following uniformly random directions throughout the entire simulation period (see \cite{banagar20193gpp}). 

Using NOMA principles, it is possible for multiple users to be supported simultaneously on each subchannel. Let $b_{k,n}(t)$ represent the subchannel assignment index in each time slot $(t)$, where $b_{k,n}(t) = 0$ denotes a vacant subchannel $k$ that can be assigned to secondary user (SU) $n$ in time slot $t$. On the other hand, if $b_{k,n}(t) = 1$
it means that the subchannel is occupied by a PU. We use $p_{k,n}(t)$ to represent the transmit power of SU $n$ in each time slot $t$. 
During the uplink transmission, each SU $n$ sends its QPSK-modulated signal to the UAV with a transmit power of $p_{k,n}$ and a channel gain of $g_{k,n}$ on subchannel $k$. QPSK modulation is a suitable choice for NOMA systems as opposed to higher-order modulation schemes because it can maintain a good level of performance while minimizing interference from other users. This is important because NOMA systems allow multiple users to share the same resources, which can lead to interference if the signals from the different users are not properly managed. The UAV is considered to fly at a constant height of $H$ above ground level, as mandated by the regulatory authority to ensure safety \cite{9120492}. To simplify the system, we assume that both the UAV and all SUs use a single antenna.
We represent the horizontal projection of the UAV trajectory as $q({t}) = [x({t}),y({t})]^T$. Hence, the time changing distance between the UAV and the ground SUs can be formulated as
\begin{equation}
    d({t}) = \sqrt{H^{2} + \lVert {q}({t})\rVert^2}, 0\leq t \leq T.
\end{equation}

Assume that the initial horizontal position of the UAV is pre-defined as $q(1) = (x_{i},y_{i})$. During each time slot $t$, the UAV can alter its position. 
For the sake of simplicity, we make the assumption that the communication channel follows a line-of-sight (LoS) path and utilize the free-space path loss (FSPL) model similar to \cite{8247211}.
As a result, the channel power gain from SU $n$  to the UAV in each time slot $t$ is represented as

\begin{equation}
    g_{k,n}(t) = \mu_{k,n}(t)\xi_{k,n}(t)\beta_{k,n}(t) d_{k,n}(t)^{-\alpha},
\end{equation}
where $\alpha$ denotes the path-loss exponent, $\mu_{k,n}(t)$ addresses the small-scale fading, $\xi_{k,n}(t)$ indicates the shadowing factor, $\beta_{k,n}(t)$ represents the reference link power gain. 
Thus, the received signal at the UAV in each time slot $t$ over $k$ is expressed as:
\begin{multline}
    y_{k}(t) = \sum^{N}_{n=1} b_{k,n}(t) g_{k,n}(t) \sqrt{p_{k,n}(t)} d_{k,n}(t) + \nu_{k}(t),
\end{multline}
where the first segment of the equation represents the transmitted signals from $n$ SUs on subchannel $k$ and $\nu_{k}(t)$ accounts for the presence of additive white Gaussian noise.
Each SU must be assigned a distinct QPSK constellation to prevent signal interference between the signals from various SUs. This ensures a minimal distance between the constellations for accurate SIC decoding at the receiver side.
%

The uplink SIC is executed in a descending order based on the channel gain. Initially, it decodes the SU with the strongest signal first, then removes it from the superposed signal, and subsequently decodes the remaining users with weaker signals. Therefore, in each time slot, the attainable data rate $\mathrm{R}_{k,n}$ of SU $n$ on sub-channel $k$ as expressed by Shannon capacity:
\begin{equation}
   \mathrm{R}_{k,n} \triangleq b_{k,n} \log_2 \bigg(1+\frac{p^{}_{k,n}g^{}_{k,n}}{\sum^{\vert\mathcal{U}_{k}\vert}_{j=\sigma^{-1}_{k}(n)+1}{p^{}_{\sigma_{k}(j)}}{g^{}_{\sigma_{k}(j)}}+{\eta_{k,n}^{}}}\bigg).
\end{equation}

The objective is to collectively optimize subchannel selection $b_{k,n}(t)$, and power allocation $p_{k,n}(t)$ to maximize the total sum rate in the cell. Mathematically, we express the problem of maximizing the sum rate as follows:
\begin{subequations}
\begin{align}\label{eq_optimizationProblem}
\begin{split}
  \max_{\{q(t), b_{k,n}(t), p_{k,n}(t)\}} \ {\sum^{K}_{k=1}} \ {\sum^{N}_{n=1}{R_{k,n}}}
  \end{split}
\\[2ex]
\text{s.t.}\qquad & \text{C1:} \quad {\sum^{{K}}_{k=1}} {b_{k,n}(t)} {p_{k,n}(t)} \leq {p}_{max}, 
  \forall {k\in \mathcal{K}},{n\in \mathcal{N}},
\\
& \text{C2:} \quad {p_{k,n}(t)}  \geq 0, \forall {k \in \mathcal{K}}, {n \in \mathcal{N}},
\\
& \text{C3:} \quad  {b_{k,n}(t)} \in\{0,1\}, \forall {k \in \mathcal{K}}, {n \in \mathcal{N}},
\\
& \text{C4:} \quad {\sum^{{N}}_{n=1}} {b_{k,n}(t)}  \leq M, k \in \mathcal{K},
\\
& \text{C5:} \quad q(1) = \hat{q_{i}} 
\\
& \text{C6:} \quad {\sum_{t=1}^{T} {\Delta}{(t)} \leq T_{max}}
\end{align}
\end{subequations}
where constraints $\text{C1}$ and $\text{C2}$ enforce that the transmit power of each SU $n$ does not exceed the maximum power limit $P_{max}$ and must be non-negative, respectively. $\text{C3}$ implies that each subchannel can be either allocated or remain unassigned. Due to SIC decoding complexity, constraint $\text{C4}$ imposes a maximum of $M$ SUs to be multiplexed per subchannel. Constraint $\text{C5}$ is the UAV’s initial position. Constraint $\text{C6}$ ensures that the UAV completes its tasks within a predefined maximum duration $T_{max}$. This is essential to avoid prolonged operations that may not be practical or feasible in real-world scenarios. 

Solving the optimization problem presented in equation \eqref{eq_optimizationProblem} is challenging due to its nonconvexity and NP-hard nature. Achieving the globally suboptimal solution requires using either random or iterative search schemes or other optimization methods. However, our work aims to solve the objective function by using a conventional optimization method (optimizer) during the offline stage. The UAV then uses the solutions provided by this method to learn a dynamic generative model that represents the wireless environment and the optimizer's decision-making processes to solve a set of training examples. Throughout the online phase, the generative model enables the UAV to anticipate the future evolution of the wireless environment, deduce the optimizer's intended actions, and rectify actions when it comes across new radio situations that may deviate from the ones it was trained on. This is especially critical for intelligent radios (i.e., UAV in our scenario), as traditional optimization methods are unsuitable for online decision-making, lack online adaptive flexibility, and require high computational resources. The following sections explore an active inference-based method to efficiently learn a representation of the wireless environment where the agent's preferences are encoded based on the solutions provided by the optimizer. This enables a mobile UAV to adapt to new situations and find an optimal subchannel assignment and power allocation policy.

\section{Proposed Method for Intelligent Trajectory and Resource Allocation }
\label{Sec_proposedMethod}
In this section, we introduce a unique representation of the optimization problem in \eqref{eq_optimizationProblem} by converting the problem into abnormality minimization. The framework employs a partially observable Markov decision process (POMDP) to describe the set of variables that constitute the optimal solution. This allows the agent to detect non-stationarity (anomalies) and adapt to new conditions by obtaining a new suboptimal model, guided by the principle of free energy minimization. 

\subsection{Problem Transformation Based on Active-GDBN}
We formalize Active-GDBN within the framework of a POMDP. At each time step $t$, the actual state of the environment $\mathrm{\tilde{S}}_{t} \in \mathbb{R}^{d_s}$ changes stochastically according to a transition function $\mathrm{\tilde{S}}_{t} \sim \mathrm{Pr}(\mathrm{\tilde{S}}_t \vert \mathrm{\tilde{S}}_{t-1},\boldsymbol{\mathcal{A}})$, which is influenced by the actions $\boldsymbol{\mathcal{A}}$ ${\in}$ $\mathbb{R}^{d_a}$ taken by the UAV. Since the actual state of the environment is typically hidden from the UAV, it can only infer through observations  $\mathrm{\tilde{Z}}_{t}$ ${\in}$ $\mathbb{R}^{d_z}$, defined as $\mathrm{\tilde{Z}}_{t}\sim\mathrm{Pr}(\mathrm{\tilde{Z}}_{t}\vert\mathrm{\tilde{S}}_{t})$. Consequently, the UAV relies on its beliefs about the actual state of the environment $\mathrm{\tilde{S}}_{t}$. 
Also, applying the Bayesian principle given a prior $\mathrm{Pr}(\mathrm{\tilde{X}}_{t} \vert \mathrm{\tilde{Z}}_{t-1})$ and likelihood $\mathrm{Pr}(\mathrm{\tilde{Z}}_{t} \vert \mathrm{\tilde{X}}_{t})$, the posterior $\mathrm{Pr}(\mathrm{\tilde{X}}_{t} \vert \mathrm{\tilde{Z}}_{t})$ can be obtained as follows \cite{chen2003bayesian}:
\begin{equation} \label{eq_BayesRule}
\begin{split}
     \mathrm{Pr}(\mathrm{\tilde{X}}_{t} \vert \mathrm{\tilde{Z}}_t) & = \frac{\mathrm{P}(\mathrm{\tilde{Z}}_{t} \vert \mathrm{\tilde{X}}_{t})\mathrm{Pr}(\mathrm{\tilde{X}}_{t} \vert \mathrm{\tilde{Z}}_{t-1})}{\mathrm{Pr}(\mathrm{\tilde{Z}}_{t} \vert \mathrm{\tilde{Z}}_{t-1})} .
\end{split}   
\end{equation}
As indicated in \eqref{eq_BayesRule}, the posterior  $\mathrm{Pr}(\mathrm{\tilde{X}}_{t} \vert \mathrm{\tilde{Z}}_{t})$, is characterized by three primary elements: the prior $\mathrm{Pr}(\mathrm{\tilde{X}}_{t} \vert \mathrm{\tilde{Z}}_{t-1})$, which represents the prior knowledge of the UAV; the likelihood $\mathrm{Pr}(\mathrm{\tilde{Z}}_{t} \vert \mathrm{\tilde{X}}_{t})$, which represents the probability of the UAV observing the evidence or data given a hidden state; the observation $\mathrm{Pr}(\mathrm{\tilde{Z}}_{t}|\mathrm{\tilde{Z}}_{t-1})$, which denotes the probability of the UAV observing the data across all possible values of the hidden states.
%

In the Active-GDBN framework, the interaction between UAV and the environment can be described as a 6-element tuple.
($\boldsymbol{\mathrm{\tilde{S}}_{t}}$, $\boldsymbol{\mathrm{\tilde{X}}_{t}}$, $\boldsymbol{\mathcal{A}}$, $\boldsymbol{\mathrm{T}_{\tau}^{pu}}$, $\boldsymbol{\Pi_{\tau}^{a}}$, $\boldsymbol{\tilde{Z}_{t}}$), where $\boldsymbol{\mathrm{\tilde{S}}_{t}}$ and $\boldsymbol{\mathrm{\tilde{X}}_{t}}$ are sets of environmental hidden states that include the SU positions, discrete subchannels, and the relative positions between the UAV and each SU. $\boldsymbol{\mathcal{A}}=\{\mathcal{A}_{}^{[\mathrm{f}]}, \mathcal{A}_{}^{[p]},\mathcal{A}_{}^{[u]}\}$ is the action space containing all the possible sub-channel decisions $b_{k,n}(t)$, power allocation $p_{k,n}(t)$, and the possible UAV's trajectories $q({t})$. 
$\boldsymbol{\mathrm{T}_{\tau}^{pu}}$ represents the dynamic transition model of PUs over time. $\boldsymbol{\Pi_{\tau}^{a^{}}}$ denotes the active inference table capturing state-action pairs, and $\boldsymbol{\tilde{\mathrm{Z}}_{t}}$ comprises a sequence of $K$ sensory signals. 
%
\subsubsection{Radio Environment Representation}
\begin{figure}[t!]
\begin{center}
\begin{minipage}[t]{.85\linewidth}
\centering
    \includegraphics[height=3.3cm]{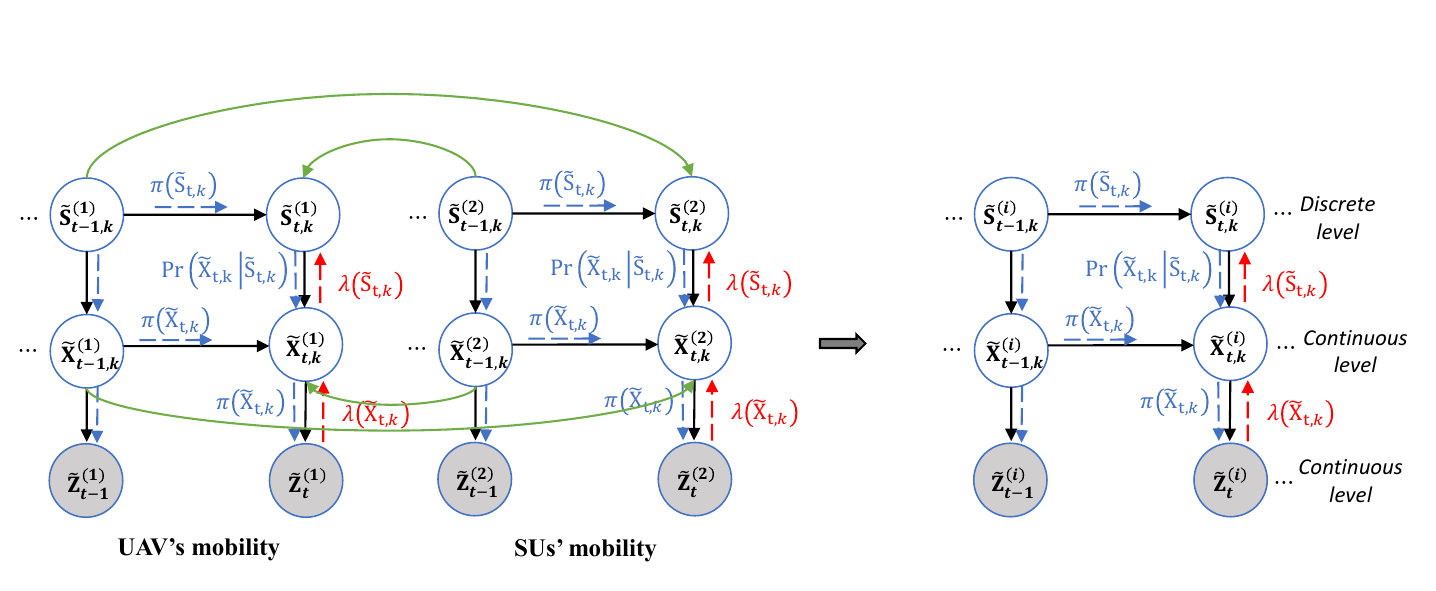}
\\[-1.0mm]
{\scriptsize (a)}
\end{minipage}
\begin{minipage}[t]{.95\linewidth}
    \centering
    \includegraphics[height=4.8cm]{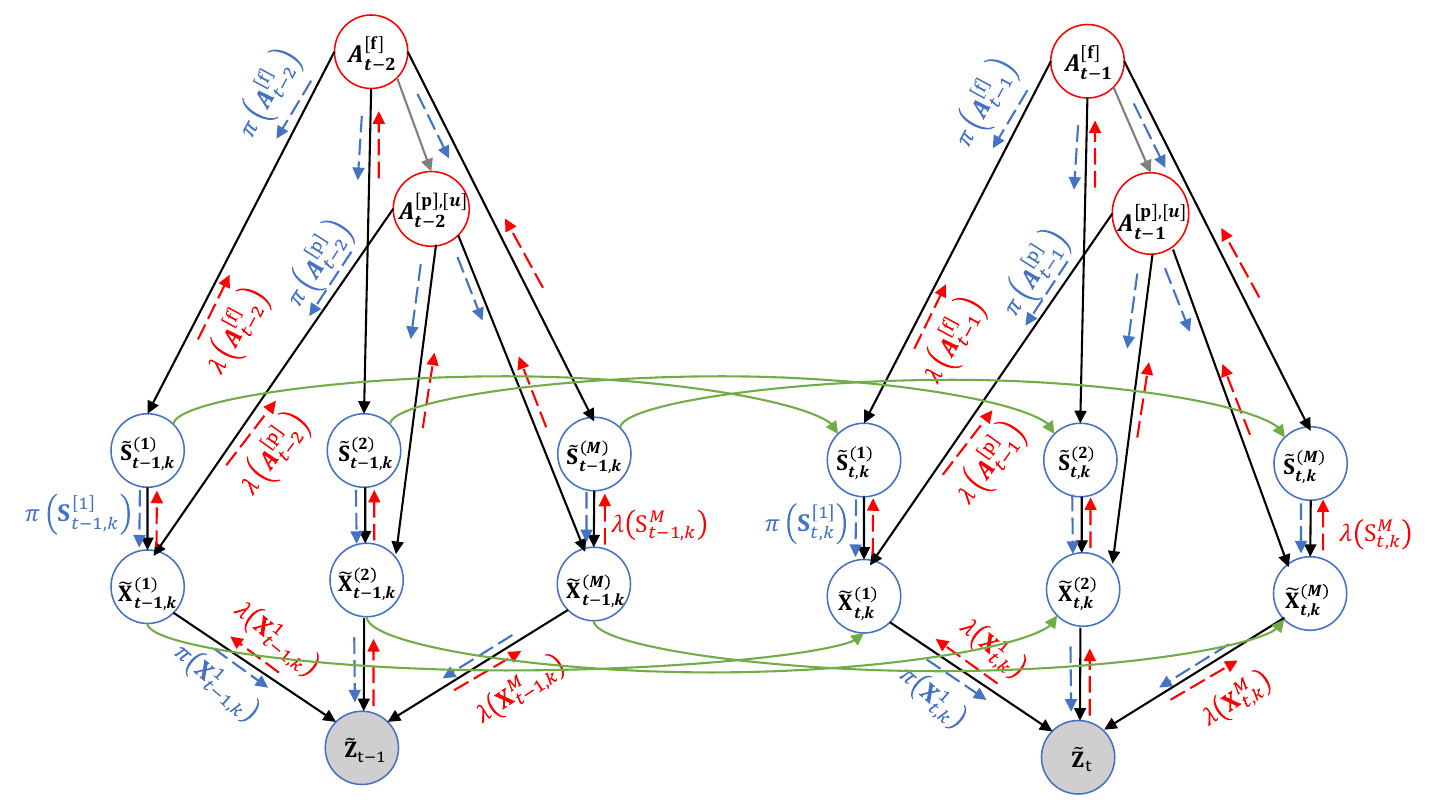}
    \\[-1.0mm]
    {\scriptsize (b)}
\end{minipage}
\end{center}
\caption{The proposed graphical representations of Active-GDBN: (a) Offline Perception (Coupled-GDBN), (b) Online Active Inference (Active-GDBN).The graph is characterized by a hierarchical structure that represents sequences of hidden states (discrete and continuous) over time. This allows the UAV to track the evolution of messages at discrete and continuous states, which it uses to estimate the posterior distribution over hidden states. The blue arrows represent messages from prior states, while the red arrows signify messages from future states. Fundamentally, the previous and future states are constantly represented, and the UAV is able to update its beliefs based on incoming new data. As displayed in sub-figure 2-b in green arrows, the current states $\mathrm{\tilde{S}}_{t,k}$, $\mathrm{\tilde{X}}_{t,k}$  at time $t$ on sub-channel $k$ determines the current observation $\mathrm{\tilde{Z}}_{t}$, which is influenced by previous states $\mathrm{\tilde{S}}_{t-1,k}$, $\mathrm{\tilde{X}}_{t-1,k}$ and previous joint actions( $A_{t-1}^{[\mathrm{f}]}$ for subchannel assignment), ( $A_{t-1}^{[\mathrm{p},{u}]}$ for continuous power allocation and UAV trajectory) .} 
    \label{fig_activeDBN}
\end{figure}

The UAV can perceive $K$ sensory signals denoted as:
$\boldsymbol{\tilde{\mathrm{Z}}_{t}} {=} \{\mathrm{\Tilde{Z}}_{t,1}, \mathrm{\Tilde{Z}}_{t,2}, \dots, \mathrm{\Tilde{Z}}_{t,K}\}$, corresponding to $K$ sub-channels. Furthermore, we use a generalized hierarchical state-space model to characterize the radio environment with the following constituents:

\begin{equation} \label{eq_discrete_model}
    \mathrm{\tilde{S}}_{t,k}^{(e)} = \textrm{f}(\mathrm{\tilde{S}}_{t-1,k}^{(e)}) + \mathrm{w}_{t,k},
\end{equation}
\begin{equation} \label{eq_continuous_model}
    \mathrm{\Tilde{X}}_{t,k}^{(e)} = \mathrm{C}\mathrm{\tilde{X}}_{t-1,k}^{(e)}+\mathrm{D}\mathrm{U}_{\mathrm{\tilde{S}}_{t,k}^{(e)}}+\mathrm{w}_{t,k},
\end{equation}
\begin{equation} \label{eq_observation_model}
    \mathrm{\Tilde{Z}}_{t,k} = \mathrm{H} \big(   \mathrm{\Tilde{X}}_{t,k}^{(1)}+\dots+\mathrm{\Tilde{X}}_{t,k}^{(M)}+\mathrm{\Tilde{X}}_{t,k}^{(pu)}\big)+\mathrm{v}_{t,k}.
\end{equation}
In equation \eqref{eq_discrete_model}, $\scriptsize \mathrm{\tilde{S}}_{t,k}^{(e)}$ represents the discrete random variables signifying discrete state clusters of the physical signal, the sub-channel transmitting the signal, and its power level.
Similarly, $\textrm{f}(.)$ denotes a nonlinear function that illustrates the evolution of $\mathrm{\tilde{S}}_{t,k}^{(e)}$ over time based on $\mathrm{\tilde{S}}_{t-1,k}^{(e)}$, and $\mathrm{w}_{t,k}$ represents the noise, given as $\mathrm{w}_{t,k} {\sim} \mathcal{N}(0, \Sigma_{\mathrm{w}_{t,k}})$. Equation \eqref{eq_continuous_model} is the dynamic model equation, which describes the temporal evolution of the Generalized States (GS) $\mathrm{\Tilde{X}}_{t,k}^{(e)}$ influenced by both $\mathrm{\Tilde{X}}_{t-1,k}^{(e)}$ and $\mathrm{\tilde{S}}_{t,k}^{(e)}$ where $e\in\{no,pu,c\}$, $no$, $pu$, and $c$ represent noise, PU and the superimposed NOMA signals, respectively. $\mathrm{C}$ and $\mathrm{D}$ are matrices that represent the dynamic and control rules, respectively, while $\mathrm{U}_{\mathrm{\tilde{S}}_{t,k}^{(e)}}$ is the control vector.
Equation
\eqref{eq_observation_model} is the observation model (sensory signals), which depends on the GS.
Fig.~\ref{fig_activeDBN} displays the proposed graphical models representing the Active-GDBN framework. 

The process includes an offline phase of perceptual learning, during which we equip the UAV with an interactive coupled-state switching GDBN at discrete and continuous levels, as shown in Fig.~\ref{fig_activeDBN}-a. This is done to effectively capture the temporal dependencies between the UAV trajectory and the SUs' mobility.  
Coupled state-switching models \cite{pohle2021primer} are designed to address multiple observation sequences, where underlying state variables interact with each other. In this context, a hidden discrete state 
$\mathrm{\tilde{S}}_{t,k}^{(1)}$ is influenced by its own previous state, $\mathrm{\tilde{S}}_{t-1,k}^{(1)}$, as well as the previous state of the other hidden chain, $\mathrm{\tilde{S}}_{t-1,k}^{(2)}$. In a similar manner, a continuous hidden state $\mathrm{\tilde{X}}_{t,k}^{(1)}$ is influenced by its own previous state, $\mathrm{\tilde{X}}_{t-1,k}^{(1)}$, and the previous state of the other hidden chain, $\mathrm{\tilde{X}}_{t-1,k}^{(2)}$. Fig.~\ref{fig_activeDBN}-b is the online active inference stage where the UAV needs to make a joint decision for all three actions,
by finding the best combination of subchannel assignment $b_{k,n}(t)$, power levels $p_{k,n}(t)$, and UAV trajectory $q({t})$  to maximize the sum rate.

\subsubsection{Offline Perceptual Learning}
At the initial learning phase, the UAV starts with an initial model identical to the Unmotivated Kalman Filter (UKF), assuming static environmental state evolution. The UAV's memory generates initial generalized errors (GEs), which are then used for incremental learning of new models.
The goal of the offline perception stage is to equip the UAV with the ability to learn different vocabularies representing the noise model, PU model, and the superimposed SUs' model. Note that the superimposed SUs' signal is generated using a matrix that encodes the relationships between all possible subchannel selection indices, power allocations, and all the possible distances between the UAV trajectory and user positions.

We consider $I$ distinct observations \{$\mathrm{\tilde{Z}}_{t}^{(i)}\}^{T}_{t=1}$, each depending on hidden environment state sequence at discrete and continuous states \{$\mathrm{\tilde{S}}_{t}^{(i)}\}^{T}_{t=1}$, \{$\mathrm{\tilde{X}}_{t}^{(i)}\}^{T}_{t=1}$, $i=1, \dots , I$. The hidden states variables interact at discrete and continuous states. The transition probabilities relating to the discrete and continuous state vectors are given by:
\begin{equation}
    \mathrm{Pr}(\mathrm{\tilde{S}}_{t} \vert \mathrm{\tilde{S}}_{t-1}) = \mathrm{Pr}(\mathrm{\tilde{S}}^{(1)}_{t}, \dots , \mathrm{\tilde{S}}^{(I)}_{t} \vert \mathrm{\tilde{S}}^{(1)}_{t-1}, \dots , \mathrm{\tilde{S}}^{(I)}_{t-1})
\end{equation}
\begin{equation}
    \mathrm{Pr}(\mathrm{\tilde{X}}_{t} \vert \mathrm{\tilde{X}}_{t-1}) = \mathrm{Pr}(\mathrm{\tilde{X}}^{(1)}_{t}, \dots , \mathrm{\tilde{X}}^{(I)}_{t} \vert \mathrm{\tilde{X}}^{(1)}_{t-1}, \dots , \mathrm{\tilde{X}}^{(I)}_{t-1})
\end{equation}
In this case $(I=2)$, the UAV's mobility and the SUs' mobility in each time step.

We applied the unsupervised clustering algorithm called Growing Neural Gas (GNG) \cite{4761768} to train the coupled GDBN model. The GNG algorithm is a self-organizing neural network that can adaptively learn the structure of the input data through cooperative and incremental learning. This model takes in the GEs and produces clusters of discrete states and a set of continuous generalized states. 

\subsubsection{Active Inference Stage}
The UAV's trajectory is divided into multiple time slots, which allows it to adapt and optimize its actions in response to changing conditions over time. 
The decision-making process of the UAV relies on the state-action pair encoded in  a time-varying matrix $\boldsymbol{\Pi_{\tau}^{[\mathrm{f}]^{}}}$, which encodes the probabilistic dependencies between states and discrete actions (subchannel assignment), while $\boldsymbol{\Pi_{\tau}^{[\mathrm{p}]^{}}}$ and $\boldsymbol{\Pi_{\tau}^{[\mathrm{u}]^{}}}$ are time-varying matrices encoding the probabilistic dependencies between states and continuous actions, power levels, and UAV trajectories, respectively.

\paragraph{Action selection}
At the beginning of each time slot, the UAV enters a specific state defined by its current position and channel conditions. By leveraging the learned vocabularies during the offline perception stage, the UAV observes the PU activities and the positions of SUs. Initially, the UAV employs random sampling to select joint actions, with each possible action having an equal chance of being selected.
The UAV selects the initial joint actions for the current time slot, which include subchannel assignment and power allocation based on the UAV's current state. 
In the subsequent iterations, the environment responds to the UAV's actions by generating new observations, updating the environmental states, and altering the positions of SUs and the UAV. As the UAV interacts with the environment it learns from the outcomes of its actions. By iteratively updating its generative model and action policies, the UAV adapts its decision-making strategies to improve its performance over time. Concurrently, the UAV can predict the future behavior of PUs using $\boldsymbol{\mathrm{T}_{\tau}^{pu}}$ and anticipate resources that are likely to be occupied by PUs.

\paragraph{Perception and joint State-Prediction}
After selecting the joint actions, which include the UAV's trajectory, subchannel assignments, and  power allocation, the UAV can correctly estimate and predict the effect of its actions by using a modified Markov Jump Particle Filter (M-MJPF) \cite{9322583}. The M-MJPF incorporates a switching model and employs particle filtering for discrete state prediction and updating, along with Kalman filtering for continuous state prediction and updating. time-dependent inter-slice top-down predictive messages $\pi(\mathrm{\tilde{X}}_{t,k})$ and $\pi(\mathrm{\tilde{S}}_{t,k})$ relies on information learned from the dynamic model. Conversely, the intra-slice bottom-up inference is established in the likelihood function and involves the transmission of backward-propagated messages $\lambda(\mathrm{\tilde{X}}_{t,k})$ and $\lambda(\mathrm{\tilde{S}}_{t,k})$  moving to the discrete level. The discrete level influences the prediction at the continuous level. For every particle propagated within the discrete level, a KF is activated to estimate the corresponding continuous level $\mathrm{\tilde{X}}_{t,k}$. The PF generates $L$ particles with equal weighting, guided by the proposal density coded in the transition matrix  $\Pi_{k}$.
\paragraph{Abnormality measurements and action evaluation}
The Bhattacharyya distance is employed in the measurement of continuous state abnormalities to compute the difference between two messages that reach node $\mathrm{\tilde{X}}_{t,{k}}^{}$, given by, $\pi(\mathrm{\tilde{X}}_{t,{k}}^{})$ and $\lambda(\mathrm{\tilde{X}}_{t,{k}}^{})$. This is done to evaluate how well the observations match the predictions made by the model, as shown below:
\begin{equation}
\scriptsize
    \boldsymbol{\Upsilon_{\mathrm{\tilde{X}}_{t,{k}}^{}}} = -\ln \bigg( \mathcal{BC}\big(\pi(\mathrm{\tilde{X}}_{t,{k}}^{}),\lambda(\mathrm{\tilde{X}}_{t,{k}}^{})\big) \bigg)= \int \sqrt{\pi(\mathrm{\tilde{X}}_{t,{k}}^{})\lambda(\mathrm{\tilde{X}}_{t,{k}}^{})} d\mathrm{\tilde{X}}_{t,{k}}^{},
    \label{CLA_abn_reference_model}
\end{equation}
where $\mathcal{BC}$ denotes the Bhattacharyya coefficient.  Bhattacharyya distance measures the divergence between two probability distributions. 
A larger distance signifies a higher difference between observations and model predictions, resulting in higher abnormalities. On the other hand, a smaller distance indicates a closer match between observations and predictions, suggesting a lower abnormality or an accurate prediction. 

\paragraph{Update of the action selection and incremental model}
\text{Action selection}: In exploitation, given the UAV’s present model and present observation, the UAV selects actions that minimize available model free energy. During exploration, the UAV selects actions to minimize the expected free energy, which is the new model resulting from the chosen action

\text{Model incremental update}: The UAV updates the POMDP for the new actions that it has learned.
%

\section{Simulation Results}
\label{Sec_simulationResults}
In this section, we provide simulation results to validate the effectiveness of the proposed Active-GDBN.

\begin{figure}[t!]
    \begin{center}
        \begin{minipage}[b]{0.70\linewidth}
        \centering
            \includegraphics[width=6.0cm]{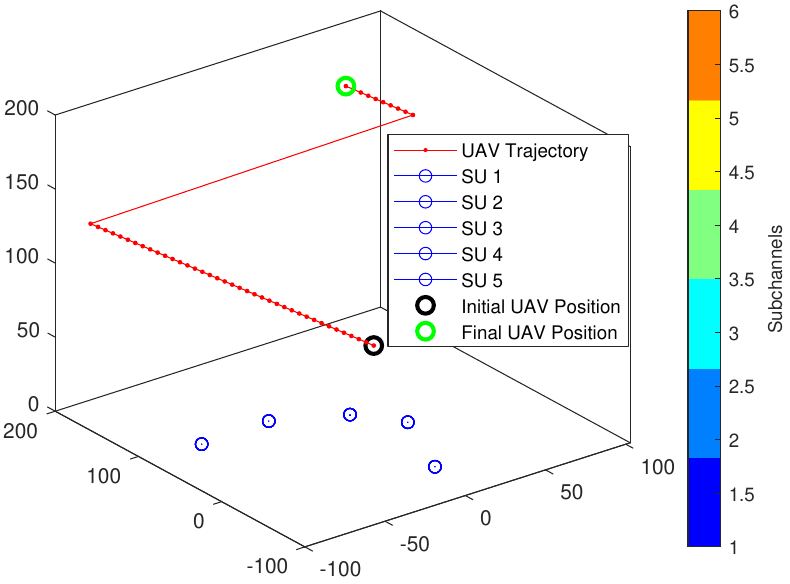}
        \\[-1.5mm]
            {\tiny (a) UAV trajectory, SU mobility, and subchannels.}
        \end{minipage}
        \\[1.0mm]
        \begin{minipage}[b]{0.70\linewidth}
            \centering
            \includegraphics[width=6.0cm]{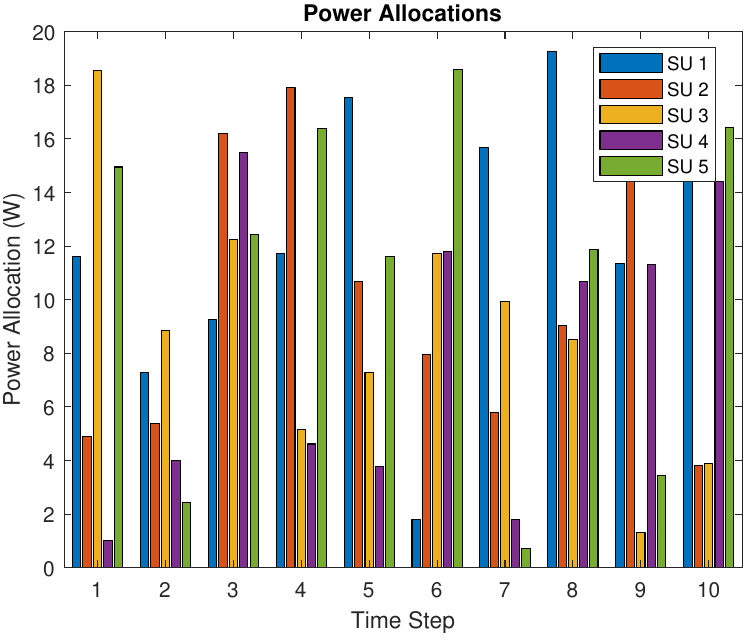}
            \\[-1.5mm]
            {\tiny (b) Power allocation in each time step}.
        \end{minipage}
        %
        \caption{Illustration of the simulation environment
        with $M = 5$, $H = 50$ m, $P_{max} = 20$ Watts.}
            \label{fig_Simulation_Setup}
    \end{center}
\end{figure}

\textbf{Data Generation}: Random initial positions for multiple SUs are generated within $0$ and $1$ for each user and their spatial 3D (x, y, and z). The user positions are updated at each time step and stored in a user mobility matrix. 
The UAV's initial position and direction are randomly generated. A loop iterates through time steps, updating the UAV's position based on its speed and direction. 
The final position is computed based on its trajectory after the specified time steps and stored in a UAV trajectory matrix. At every time step, power allocations are randomly generated within the specified maximum power $(P_{max})$. These are stored in a power allocation  matrix, where each row represents a user and each column represents a time step. The subchannel assignment matrix is populated with random integers within the range $1$ to $6$ for all possible subchannels. We then generate a distance matrix encoding all possible distances between the UAV trajectory and user positions. 
%
Table I presents an overview of the network parameters, while for illustrative purposes, Figs.~\ref{fig_Simulation_Setup}-a  and  ~\ref{fig_Simulation_Setup}-b depict the simulation setup.

\begin{table}[!ht]
\begin{center}
\caption{Simulation Parameters}
\scalebox{0.75}{%
\begin{tabular}{ |l||l| }
 \hline
 Cell radius & 100 m \\ 
 \hline
 UAV flight height $(H)$ & 50 m \\ 
 \hline
 UAV speed  & 5 m/s \\ 
 \hline
 Number of UAV time steps & 10 \\ 
 \hline
 Modulation scheme of PUs & BPSK  \\ 
 \hline
 Modulation scheme of SUs & QPSK  \\ 
 \hline
 Path loss model & Free-space-path-loss \cite{wu2018joint}  \\ 
 \hline
 Noise power & $-174$ dBm/Hz  \\ 
 \hline
 System Bandwidth, $B_{w}$ & 1.4 MHz  \\ 
 \hline
 Number of sub-channels  $K$ & 6  \\ 
 \hline
 Power budget of SUs $P_{max}$ &  20 W\cite{8247211} \\ 
\hline
 Maximum number of SUs per sub-channel $M$ & $M = [1,2,3,4,5,6]$,  \\ 
 \hline
 Power difference threshold $P_{th}$ & $1$  \\ 
 \hline
 Learning rate of GNG clustering & $0.01$  \\ 
 \hline
\end{tabular}}
\end{center}
\end{table}

\begin{figure}[ht!]
    \centering
    \hspace*{-0.5cm}
    \includegraphics[height=3.0cm]{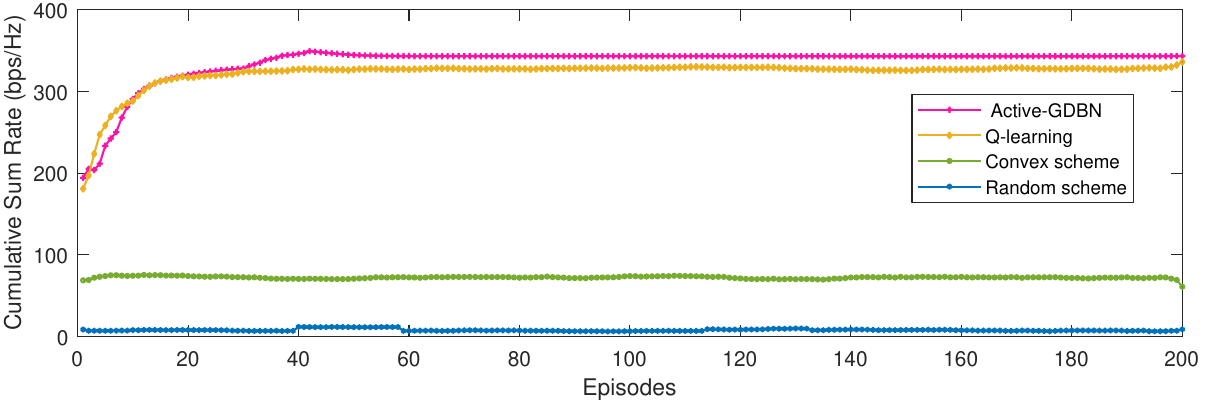}
    \caption{Cumulative sum rate comparison of the proposed Active-GDBN with benchmark schemes when $M = 2$, $H = 50$ m, $P_{max} = 20$ Watts.}
    \label{fig_variable_benchmark}
\end{figure}

To assess the effectiveness of the proposed Active-GDBN, we benchmark with suboptimal baselines such as q-learning \cite{9120492}, convex \cite{wu2018joint}, and random schemes.
As evident from Fig.~\ref{fig_variable_benchmark}, the proposed approach performs better than the benchmark schemes by achieving an improved and consistent sum rate in fewer episodes. This is because Active-GDBN inherently balances exploration and exploitation by seeking to minimize abnormalities or prediction errors in each episode to update its beliefs. The UAV continuously updates its internal model online to align with new sensory inputs. On the other hand, Q-learning necessitates manual adjustment of exploration strategies and extensive trial and error to achieve enhanced sum rate performance. The low sum rate performance of convex and random schemes is due to a lack of inherent adaptability and online learning to adapt to time-varying environments.

\begin{figure}[ht!]
    \centering
    \hspace*{-0.5cm}
    \includegraphics[height=3.0cm]{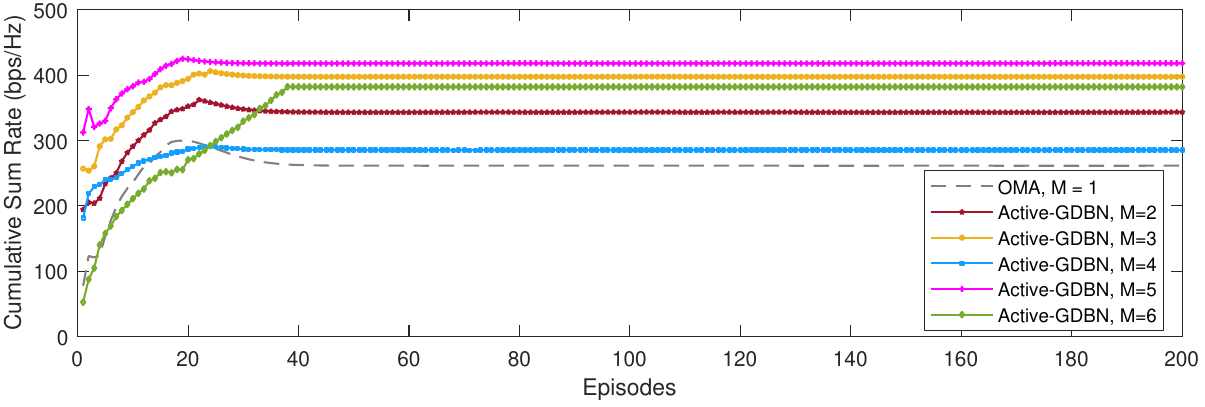}
    \caption{Convergence of Active-GDBN with varying numbers of multiplexed SUs $M$, $H = 50$ m, $P_{max} = 20$ Watts.}
    \label{fig_variable_M}
\end{figure}

Fig.~\ref{fig_variable_M} demonstrates the convergence performance of Active-GDBN using a varying number of multiplexed SUs per subchannel ($M$). When the value of $M$ is $1$, the situation becomes orthogonal multiple access (OMA). We observed that stable convergence is achieved within a range of zero to forty episodes. As anticipated, the sum rate for each SU increases monotonically, and the proposed Active-GDBN achieved a maximum of $5$ SUs per subchannel. However, this monotonic behavior is not guaranteed when $M = 4$. This deviation could be attributed to the presence of reduced LoS conditions, arising from the UAV's random trajectory. This randomness introduces unexpected interference for a specific $M$ configuration, resulting in a drop in the cumulative sum rate. Similarly, the algorithm exhibits performance degradation as we increase the value of $M$ to $6$. This is because, at a certain limit, due to power control, simultaneous transmission from superimposed SUs begins to overlap, leading to a decrease in the sum rate.

\begin{figure}[ht!]
    \centering
    \hspace*{-0.5cm}
    \includegraphics[height=3.0cm]{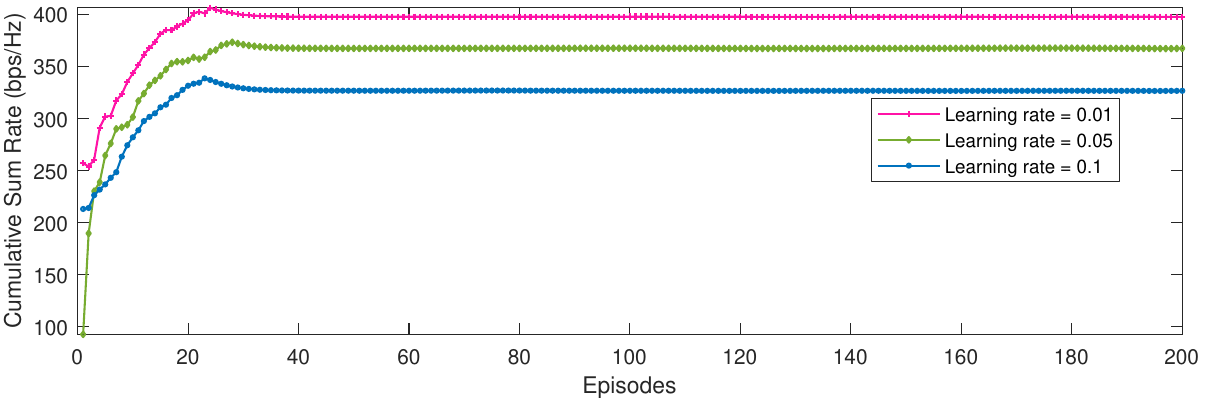}
    \caption{Effect of the GNG learning rate on convergence during offline perception training when $M = 3$, $H = 50$ m, $P_{max} = 20$ Watts.}
    \label{fig_variable_Lrate}
\end{figure}

The GNG learning rate determines the speed at which the UAV adjusts to the sensory data. As clearly indicated in Fig.~\ref{fig_variable_Lrate}, we fix the learning rate at 0.01 for all simulation settings to attain a faster and enhanced sum rate.

\section{Conclusion} \label{Sec_conclusion}
In this study, we present a novel framework based on active inference for maximizing the cumulative sum rate of a UAV-based cognitive NOMA network. By leveraging active inference, we introduce a more robust approach that can handle the complexities of dynamic wireless networks. The UAV constantly updates a generative model online to optimize subchannel assignment, power allocation, and UAV trajectories. Simulation results have shown that the proposed Active-GDBN can achieve an improved cumulative sum rate compared to suboptimal baseline techniques. Future research will explore the effects of incorporating multiple UAVs, numerous ground users, and conducting real-world experiments.
\bibliographystyle{IEEEtran}
\bibliography{References}

\begin{thebibliography}{10}
\providecommand{\url}[1]{#1}
\csname url@samestyle\endcsname
\providecommand{\newblock}{\relax}
\providecommand{\bibinfo}[2]{#2}
\providecommand{\BIBentrySTDinterwordspacing}{\spaceskip=0pt\relax}
\providecommand{\BIBentryALTinterwordstretchfactor}{4}
\providecommand{\BIBentryALTinterwordspacing}{\spaceskip=\fontdimen2\font plus
\BIBentryALTinterwordstretchfactor\fontdimen3\font minus \fontdimen4\font\relax}
\providecommand{\BIBforeignlanguage}[2]{{%
\expandafter\ifx\csname l@#1\endcsname\relax
\typeout{** WARNING: IEEEtran.bst: No hyphenation pattern has been}%
\typeout{** loaded for the language `#1'. Using the pattern for}%
\typeout{** the default language instead.}%
\else
\language=\csname l@#1\endcsname
\fi
#2}}
\providecommand{\BIBdecl}{\relax}
\BIBdecl

\bibitem{elmokadem2021towards}
T.~Elmokadem and A.~V. Savkin, ``Towards fully autonomous uavs: A survey,'' \emph{Sensors}, vol.~21, no.~18, p. 6223, 2021.

\bibitem{9120492}
Y.~Huang, X.~Mo, J.~Xu, L.~Qiu, and Y.~Zeng, ``Online maneuver design for uav-enabled noma systems via reinforcement learning,'' in \emph{2020 IEEE Wireless Communications and Networking Conference (WCNC)}, 2020, pp. 1--6.

\bibitem{he2019joint}
C.~He, Y.~Hu, Y.~Chen, and B.~Zeng, ``Joint power allocation and channel assignment for noma with deep reinforcement learning,'' \emph{IEEE Journal on Selected Areas in Communications}, vol.~37, no.~10, pp. 2200--2210, 2019.

\bibitem{salaun2018optimal}
L.~Sala{\"u}n, C.~S. Chen, and M.~Coupechoux, ``Optimal joint subcarrier and power allocation in noma is strongly np-hard,'' in \emph{2018 IEEE International Conference on Communications (ICC)}.\hskip 1em plus 0.5em minus 0.4em\relax IEEE, 2018, pp. 1--7.

\bibitem{kurenkov2018reinforcement}
A.~Kurenkov, ``Reinforcement learning’s foundational flaw,'' \emph{The Gradient}, 2018.

\bibitem{ccatal2020learning}
O.~{\c{C}}atal, S.~Wauthier, C.~De~Boom, T.~Verbelen, and B.~Dhoedt, ``Learning generative state space models for active inference,'' \emph{Frontiers in Computational Neuroscience}, vol.~14, p. 574372, 2020.

\bibitem{friston2017graphical}
K.~J. Friston, T.~Parr, and B.~de~Vries, ``The graphical brain: Belief propagation and active inference,'' \emph{Network neuroscience}, vol.~1, no.~4, pp. 381--414, 2017.

\bibitem{sajid2022active}
N.~Sajid, P.~Tigas, and K.~Friston, ``Active inference, preference learning and adaptive behaviour,'' in \emph{IOP Conference Series: Materials Science and Engineering}, vol. 1261, no.~1.\hskip 1em plus 0.5em minus 0.4em\relax IOP Publishing, 2022, p. 012020.

\bibitem{da2020active}
L.~Da~Costa, T.~Parr, N.~Sajid, S.~Veselic, V.~Neacsu, and K.~Friston, ``Active inference on discrete state-spaces: A synthesis,'' \emph{Journal of Mathematical Psychology}, vol.~99, p. 102447, 2020.

\bibitem{9829873}
A.~Krayani, A.~S. Alam, L.~Marcenaro, A.~Nallanathan, and C.~Regazzoni, ``{A Novel Resource Allocation for Anti-Jamming in Cognitive-UAVs: An Active Inference Approach},'' \emph{IEEE Communications Letters}, vol.~26, no.~10, pp. 2272--2276, Oct 2022.

\bibitem{wu2018joint}
Q.~Wu, Y.~Zeng, and R.~Zhang, ``Joint trajectory and communication design for multi-uav enabled wireless networks,'' \emph{IEEE Transactions on Wireless Communications}, vol.~17, no.~3, pp. 2109--2121, 2018.

\bibitem{li2019fundamental}
P.~Li and J.~Xu, ``Fundamental rate limits of uav-enabled multiple access channel with trajectory optimization,'' \emph{IEEE Transactions on Wireless Communications}, vol.~19, no.~1, pp. 458--474, 2019.

\bibitem{al2014optimal}
A.~Al-Hourani, S.~Kandeepan, and S.~Lardner, ``Optimal lap altitude for maximum coverage,'' \emph{IEEE Wireless Communications Letters}, vol.~3, no.~6, pp. 569--572, 2014.

\bibitem{lyu2016placement}
J.~Lyu, Y.~Zeng, R.~Zhang, and T.~J. Lim, ``Placement optimization of uav-mounted mobile base stations,'' \emph{IEEE Communications Letters}, vol.~21, no.~3, pp. 604--607, 2016.

\bibitem{zeng2016throughput}
Y.~Zeng, R.~Zhang, and T.~J. Lim, ``Throughput maximization for uav-enabled mobile relaying systems,'' \emph{IEEE Transactions on communications}, vol.~64, no.~12, pp. 4983--4996, 2016.

\bibitem{bayerlein2018trajectory}
H.~Bayerlein, P.~De~Kerret, and D.~Gesbert, ``Trajectory optimization for autonomous flying base station via reinforcement learning,'' in \emph{2018 IEEE 19th International Workshop on Signal Processing Advances in Wireless Communications (SPAWC)}.\hskip 1em plus 0.5em minus 0.4em\relax IEEE, 2018, pp. 1--5.

\bibitem{challita2019interference}
U.~Challita, W.~Saad, and C.~Bettstetter, ``Interference management for cellular-connected uavs: A deep reinforcement learning approach,'' \emph{IEEE Transactions on Wireless Communications}, vol.~18, no.~4, pp. 2125--2140, 2019.

\bibitem{zeng2019path}
Y.~Zeng and X.~Xu, ``Path design for cellular-connected uav with reinforcement learning,'' in \emph{2019 IEEE Global Communications Conference (GLOBECOM)}.\hskip 1em plus 0.5em minus 0.4em\relax IEEE, 2019, pp. 1--6.

\bibitem{zeng2017energy}
Y.~Zeng and R.~Zhang, ``Energy-efficient uav communication with trajectory optimization,'' \emph{IEEE Transactions on wireless communications}, vol.~16, no.~6, pp. 3747--3760, 2017.

\bibitem{banagar20193gpp}
M.~Banagar and H.~S. Dhillon, ``3gpp-inspired stochastic geometry-based mobility model for a drone cellular network,'' in \emph{2019 IEEE Global Communications Conference (GLOBECOM)}.\hskip 1em plus 0.5em minus 0.4em\relax IEEE, 2019, pp. 1--6.

\bibitem{8247211}
Q.~Wu, Y.~Zeng, and R.~Zhang, ``Joint trajectory and communication design for multi-uav enabled wireless networks,'' \emph{IEEE Transactions on Wireless Communications}, vol.~17, no.~3, pp. 2109--2121, 2018.

\bibitem{chen2003bayesian}
Z.~Chen \emph{et~al.}, ``Bayesian filtering: From kalman filters to particle filters, and beyond,'' \emph{Statistics}, vol. 182, no.~1, pp. 1--69, 2003.

\bibitem{pohle2021primer}
J.~Pohle, R.~Langrock, M.~v.~d. Schaar, R.~King, and F.~H. Jensen, ``A primer on coupled state-switching models for multiple interacting time series,'' \emph{Statistical Modelling}, vol.~21, no.~3, pp. 264--285, 2021.

\bibitem{4761768}
I.~J. Sledge and J.~M. Keller, ``Growing neural gas for temporal clustering,'' in \emph{2008 19th International Conference on Pattern Recognition}, 2008, pp. 1--4.

\bibitem{9322583}
A.~Krayani, M.~Baydoun, L.~Marcenaro, A.~S. Alam, and C.~Regazzoni, ``Self-learning bayesian generative models for jammer detection in cognitive-uav-radios,'' in \emph{GLOBECOM 2020 - 2020 IEEE Global Communications Conference}, 2020, pp. 1--7.

\end{thebibliography}

\vspace{12pt}

\end{document}